\documentclass{article}
\usepackage{amsfonts}
\usepackage{amsmath}

\begin{document}

\title{On algebraic models of relativistic scattering}
\author{G. A. Kerimov$^{1}$ and A. Ventura$^{2,3}$ \\
$^{1}$\textit{Physics Department, Trakya University, 22030 Edirne, Turkey}\\
$^{2}$\textit{Ente per le Nuove Tecnologie, l'Energia e l'Ambiente, Bologna,
Italy}\\
$^{3}$\textit{Istituto Nazionale di Fisica Nucleare, Sezione di Bologna,
Italy}}
\maketitle

\begin{abstract}
In this paper we develop an algebraic technique for building relativistic
models in the framework of the direct-interaction theories. The interacting
mass operator $M$ in the Bakamjian-Thomas construction is related to a
quadratic Casimir operator $C$\ of non-compact group $G$.\ As a consequence
the $S$ matrix can be gained from an intertwining relation between
Weyl-equivalent representation of $G$. The method is illustrated by explicit
application to a model with $SO(3,1)$ dynamical symmetry.
\end{abstract}

\section{Introduction}

The role of dynamical symmetries in quantum theory has been emphasized by
many authors and its implications for physically relevant quantities have
been widely recognized. In this connection the importance and relevance of
dynamical symmetries has been discussed in several directions \cite{Ref1,
FVI94, IL95}.

Since the work of Zwanziger \cite{Zw67} it has become clear that
group-theoretical methods can be successfully applied to the solution of
scattering problems. In that paper Zwanziger has shown how the symmetry
group $SO(3,1)$ allows for an algebraic determination of the Coulomb
S-matrix elements. A fundamental step towards a deep understanding of
scattering problems in a group-theoretical framework was made by the Yale
group and others \cite{Ref2} with their method of Euclidean connection. The
key point in this major development lies in the observation that the
dynamical group $G$ that describes the scattering system in presence of
interactions can be obtained by deformation \cite{Gi74} of the group $G_{0}$
(called asymptotic group) describing the system in absence of interactions.
It appears that knowledge of the interrelation between the representations
of $G$ with those of $G_{0}$ allows purely algebraic calculations of
S-matrix elements for systems whose Hamiltonian $H$\ ( in the centre of mass
system) belongs to the centre of the enveloping algebra of \ $G$, \textit{%
i.e.}, 
\begin{equation}
H=f\left( C\right)
\end{equation}%
where $C$\ is the Casimir operator of $G$. Later on \cite{Ke98}, it has been
argued that the $S$ matrix for systems under consideration is associated
with intertwining operators between Weyl equivalent representations of \ $G$
(see below). At this stage we note that the operator $A$\ is said to
intertwine the representation $T^{\chi }$ and $T^{\tilde{\chi}}$ of the
group $G$ if the relation%
\begin{equation}
AT^{\chi }\left( g\right) =T^{\tilde{\chi}}\left( g\right) A\;,\qquad \text{%
for\ all}\;g\in G
\end{equation}%
or equivalently, 
\begin{equation}
AdT^{\chi }\left( a\right) =dT^{\tilde{\chi}}\left( a\right) A,\qquad \text{%
for\ all}\;a\in \mathfrak{g}
\end{equation}%
holds where $dT^{\chi }$ and $dT^{\tilde{\chi}}$ are the corresponding
representations of the algebra $\mathfrak{g}$ of $G$ \cite{KS67, Sc71}.

The hypothesis that scattering systems can be completely described by some
dynamical group has been verified for almost all interesting
non-relativistic problems. Moreover, the algebraic approach is useful not
only for systems with exact symmetry, but also for systems with broken
symmetry. In this case the arguments in an expression of the $S$ matrix with
an exact symmetry are substituted by generic functions of scattering
variables, called algebraic potentials \cite{Ref3, Ref4}.

Contrary to the non-relativistic case, the group-theoretical approach to
relativistic scattering has not been exploited yet, with the main exception
of scattering of a Dirac particle in a Coulomb potential \cite{Ref5}, or
Coulomb plus scalar potentials \cite{VS02}. In their study the authors use
relativistic wave equations. The algebraic approach, however, is more
general, since it relies on a symmetry and does not make any explicit
reference to an equation of motion.

Interestingly, there exists an alternative approach to the relativistic
particle dynamics based on the work of Bakamjian and Thomas \cite{BT53}
which has the advantage of being somewhat group-theoretical. The point is
that one can consider the problem of construction of relativistic theories
as that of construction of unitary representations of the inhomogeneous
Lorentz group, $ISO(3,1)$, also known as Poincar\'{e} group $\mathcal{P}$ 
\cite{Wi39}.

The Lie algebra of the Poincar\'{e} group has ten basis elements, which can
be chosen as $H$, $\mathbf{P}$, $\mathbf{J}$ and $\mathbf{K}$, which are the
generators of time translations, space translations, space rotations, and
pure Lorentz transformations, respectively. They satisfy the commutation
relations 
\begin{align}
\left[ P_{i},P_{j}\right] & =0,\quad \left[ P_{i},H\right] =0,\quad \left[
J_{i},H\right] =0  \label{P_1} \\
\left[ J_{i},J_{j}\right] & =i\epsilon _{ijk}J_{k},\quad \left[ K_{i},K_{j}%
\right] =-i\epsilon _{ijk}J_{k},\quad \left[ J_{i},K_{j}\right] =i\epsilon
_{ijk}K_{k}  \label{P_2} \\
\left[ J_{i},P_{j}\right] & =i\epsilon _{ijk}P_{k},\quad \left[ P_{i},K_{j}%
\right] =-i\delta _{ij}H.  \label{P_3}
\end{align}%
( Throughout this paper units are used in which $\hbar =c=1$.) Here $\delta
_{ij}$\ is the Kronecker symbol, $\epsilon _{ijk}$\ the Levi-Civita symbol
and the summation convention on repeated indices is assumed. The operators $%
H,\mathbf{P}$\ and $\mathbf{J}$\ have the physical significance of energy,
momentum, and angular momentum. According to Ref.\cite{Wi39} an elementary
particle should be described by positive energy unitary irreducible
representation $(m,s,+)$\ of the Poincar\'{e} group $\mathcal{P}$, where $m$
is the mass and $s$ denotes the spin and $+$\ means positive energy.
Therefore, the description of $N$\ noninteracting particles is given by the
tensor product of representations $(m_{i},s_{i},+),$\ $i=1,2,\ldots $.

The problem of adding interactions to the noninteracting representation of
the Poincar\'{e} group $\mathcal{P}$ consistent with the commutation
relations (\ref{P_1}) has been discussed by Dirac \cite{Di49}. Although
Dirac did not propose a practical method of constructing an interacting
representation of the Poincar\'{e} group $\mathcal{P}$, he emphasized that
there are three possible schemes for incorporating interactions into the
noninteracting representation. These schemes are now called
\textquotedblright instant form\textquotedblright , \textquotedblright front
form\textquotedblright\ and \textquotedblright point form\textquotedblright
. Later on, Bakamjian and Thomas \cite{BT53} have proposed a method for
adding interactions to a noninteracting representation of the Poincar\'{e}
group. In their approach a set of 10 auxiliary operators is introduced that
satisfies simple commutation relations. For example, in the instant form the
10 auxiliary operators are $\left\{ \mathbf{P,S,X},M\right\} $ with
commutation relations 
\begin{equation}
\left[ P_{i},X_{j}\right] =-i\delta _{ij},\qquad \left[ S_{i},S_{j}\right]
=i\epsilon _{ijk}S_{k}
\end{equation}%
all other commutators vanishing, where $\mathbf{S}$\ is the intrinsic spin, $%
\mathbf{X=}i\mathbf{\nabla }_{P}$, and $M$\ is the invariant mass operator.
The Poincar\'{e} generators are then expressed in terms of $\left\{ \mathbf{%
P,S,X},M\right\} $\ according to 
\begin{equation}
H=\sqrt{M^{2}+\mathbf{P}^{2}},\qquad \mathbf{J}=\mathbf{X\times P+S,\qquad K=%
}\frac{1}{2}\left( \mathbf{XH+HX}\right) +\frac{\mathbf{P\times S}}{H+M}
\end{equation}%
\ In the Bakamjian-Thomas approach interactions are added to the mass
operator $M$, while leaving the other nine operators equal to those of the
noninteracting system (for review see Ref. \cite{KP91}). As a result, the
problem is reduced to an eigenvalue equation for the mass operator $M$.

The question naturally arises: can a group structure be introduced into the
space on which the mass operator $M$ is defined, and if so, does it also
have useful consequences? Here the mass operator $M$ is assumed to be a
function of the Casimir operator of a non-compact group. This allows pure
group-theoretical description of the $S$\ matrix. We apply this construction
to a scattering system with $SO(3,1)$\ dynamical symmetry.

\section{Two-body systems}

We consider a system of two interacting spinless particles of mass $m_{1}$\
and $m_{2}$. In building up a relevant representation of the Poincar\'{e}
group $\mathcal{P}$, it is convenient to start with the free system 
\begin{align}
H_{0}& =\sum_{a=1}^{2}h_{a},\qquad \mathbf{\hat{P}}_{0}=\sum_{a=1}^{2}%
\mathbf{\hat{p}}_{a},\qquad \mathbf{J}_{0}=\sum_{a=1}^{2}\left( \mathbf{\hat{%
x}}_{a}\times \mathbf{\hat{p}}_{a}\right) \\
\mathbf{K}_{0}& =\sum_{a=1}^{2}\frac{1}{2}\left( \mathbf{\hat{x}}%
_{a}h_{a}+h_{a}\mathbf{\hat{x}}_{a}\right) ,
\end{align}%
where 
\begin{equation}
h_{a}=\sqrt{m_{a}^{2}+\mathbf{\hat{p}}_{a}^{2}}.
\end{equation}%
The operator $\mathbf{\hat{x}}_{a}$\ is canonically conjugate to $\mathbf{%
\hat{p}}_{a}$%
\begin{equation}
\left[ \hat{p}_{a}^{i},\hat{x}_{a}^{j}\right] =-i\delta ^{ij}
\end{equation}

The non-interacting mass operator $\hat{M}_{0}$\ is defined by 
\begin{equation}
\hat{M}_{0}^{2}=H_{0}^{2}-\mathbf{\hat{P}}_{0}^{2}
\end{equation}%
It commutes with all generators of the Poincar\'{e} group $\mathcal{P}$.

The basis states of the carrier space of this representation can be taken as
the tensor products of the single-particle states. They are defined by 
\begin{equation}
\mathbf{\hat{p}}_{a}\left\vert \mathbf{p}_{1}\mathbf{p}_{2}\right\rangle
\equiv \mathbf{p}_{a}\left\vert \mathbf{p}_{1}\mathbf{p}_{2}\right\rangle
,\qquad a=1,2
\end{equation}
and normalized so that 
\begin{equation}
\left\langle \mathbf{p}_{1}\mathbf{p}_{2}\right\vert \left. \mathbf{p}%
_{1}^{\prime }\mathbf{p}_{2}^{\prime }\right\rangle =\delta ^{3}\left( 
\mathbf{p}_{1}-\mathbf{p}_{1}^{\prime }\right) \delta ^{3}\left( \mathbf{p}%
_{2}-\mathbf{p}_{2}^{\prime }\right)
\end{equation}
and 
\begin{equation}
\int d\mathbf{p}_{1}d\mathbf{p}_{2}\left\vert \mathbf{p}_{1}\mathbf{p}%
_{2}\right\rangle \left\langle \mathbf{p}_{1}\mathbf{p}_{2}\right\vert =1
\end{equation}
It will be convenient to make a change of variables from $\mathbf{p}_{1}$and 
$\mathbf{p}_{2}$\ to $\mathbf{P}$\ and $\mathbf{k}$, with $\mathbf{P}$\ the
total momentum and $\mathbf{k}$ the relative momentum. These variables are
related to $\mathbf{p}_{1}$ and $\mathbf{p}_{2}$\ by the equations \cite%
{FS64} 
\begin{equation}
\mathbf{P=p}_{1}+\mathbf{p}_{2},\qquad \mathbf{k=}\left( \varepsilon _{2}%
\mathbf{p}_{1}-\varepsilon _{1}\mathbf{p}_{2}\right) /\left( \varepsilon
_{1}+\varepsilon _{2}\right)
\end{equation}
with 
\begin{equation}
\varepsilon _{a}=\frac{1}{2}\left[ E_{a}+w_{a}\right]
\end{equation}
where $E_{a}$\ and $w_{a}$\ are given by 
\begin{equation}
E_{a}=E_{a}\left( \mathbf{p}_{a}\right) =\sqrt{m_{a}^{2}+\mathbf{p}_{a}^{2}}%
,\qquad w_{a}=w_{a}\left( \mathbf{k}\right) =\sqrt{m_{a}^{2}+\mathbf{k}^{2}}
\end{equation}
Note that the relative momentum $\mathbf{k}$\ is equal to the three-momentum
of particle 1\ in the center of mass system ( $\mathbf{P}=0$ ). Hence, $%
w_{a} $ is the c.m. energy of a particle of mass $m_{a}$. The total c.m.
energy can be expressed in the Poincar\'{e}-invariant form 
\begin{equation}
w_{1}\left( \mathbf{k}\right) +w_{2}\left( \mathbf{k}\right) =\sqrt{s}
\end{equation}
where 
\begin{equation}
s=(p_{1}+p_{2})^{2}=\left( E_{1}+E_{2}\right) ^{2}-\left( \mathbf{p}_{1}+%
\mathbf{p}_{2}\right) ^{2}.
\end{equation}

The state vectors $\left| \mathbf{Pk}\right\rangle $ and $\left| \mathbf{p}%
_{1}\mathbf{p}_{2}\right\rangle $\ are related to each other via \cite{FS64} 
\begin{equation}
\left| \mathbf{Pk}\right\rangle =\left[ J\left( \mathbf{p}_{1},\mathbf{p}%
_{2}\right) \right] ^{\frac{1}{2}}\left| \mathbf{p}_{1}\mathbf{p}%
_{2}\right\rangle
\end{equation}
where the Jacobian $J\left( \mathbf{p}_{1},\mathbf{p}_{2}\right) $ is given
by 
\begin{equation}
J\left( \mathbf{p}_{1},\mathbf{p}_{2}\right) =\left| \frac{\partial \left( 
\mathbf{p}_{1}\mathbf{p}_{2}\right) }{\partial \left( \mathbf{P,k}\right) }%
\right| =\frac{E_{1}E_{2}}{E_{1}+E_{2}}\dfrac{w_{1}+w_{2}}{w_{1}w_{2}}
\end{equation}
In the basis $\left\{ \left| \mathbf{Pk}\right\rangle \right\} $, the
non-interacting Hamiltonian $H_{0}$\ and the non-interacting mass operator $%
\hat{M}_{0}$\ are multiplication operators 
\begin{equation}
H_{0}\left| \mathbf{Pk}\right\rangle =\left( w^{2}+\mathbf{P}^{2}\right)
^{1/2}\left| \mathbf{Pk}\right\rangle ,\qquad \hat{M}_{0}\left| \mathbf{Pk}%
\right\rangle =w\left| \mathbf{Pk}\right\rangle ,
\end{equation}
where 
\begin{equation}
w=w\left( \mathbf{k}\right) =\sqrt{m_{1}^{2}+\mathbf{k}^{2}}+\sqrt{m_{2}^{2}+%
\mathbf{k}^{2}}.  \label{cme}
\end{equation}
Other generators are 
\begin{equation}
\mathbf{J}_{0}=\mathbf{\hat{X}\times \hat{P}+\hat{l},\qquad K}_{0}=\frac{1}{2%
}\left( \mathbf{\hat{X}}H_{0}\mathbf{+\hat{X}}H_{0}\right) -\frac{\mathbf{%
\hat{l}\times \hat{P}}}{M_{0}+H_{0}}
\end{equation}
where 
\begin{equation}
\mathbf{\hat{l}=\hat{\rho}\times \hat{k}}  \label{angl}
\end{equation}
is the internal angular momentum operator. The operators $\mathbf{\hat{\rho}}
$\ and $\mathbf{\hat{X}}$\ are canonically conjugate to $\mathbf{\hat{k}}$\
and $\mathbf{\hat{P}}$ and therefore 
\begin{equation}
\left[ \hat{k}_{i},\hat{\rho}_{j}\right] =-i\delta _{ij},  \label{ccr_1}
\end{equation}
and 
\begin{equation}
\left[ \hat{P}_{i},\hat{X}_{j}\right] =-i\delta _{ij}.  \label{ccr_2}
\end{equation}

To introduce the interaction one lets $M_{0}\rightarrow M$\ , where the
interacting mass operator $\hat{M}$\ is assumed to be the sum of the
non-interacting mass operator $\hat{M}_{0}$\ plus the mass-operator
interaction $\hat{V}$%
\begin{equation}
\hat{M}=\hat{M}_{0}+\hat{V}.
\end{equation}%
The set of operators $H$, $\mathbf{P}$, $\mathbf{J}$ and $\mathbf{K}$\ will
satisfy the commutation relations of the Poincar\'{e} group $\mathcal{P}$
provided that the following conditions for $\hat{V}$\ are satisfied 
\begin{equation}
\left[ \hat{V},\mathbf{P}\right] =0,\qquad \left[ \hat{V},\mathbf{J}\right]
=0
\end{equation}%
(These constraints lead to conservation of linear and angular momenta for
the interacting system.) In \cite{BT53} the operator $\hat{V}$\ is taken to
be a ( rotationally ) scalar operator function of $\mathbf{\hat{k}}$\textbf{%
\ }and $\mathbf{\hat{\rho}}$ only 
\begin{equation}
\hat{V}=V\left( \mathbf{\hat{k},\hat{\rho}}\right)
\end{equation}

The scattering theory within the framework of the Bakamjian-Thomas
construction has been considered by the several authors \cite{KP91, FS64,
JMS64, Co65, CP82, Fu01}. The \textquotedblleft in\textquotedblright\ and
\textquotedblleft out\textquotedblright\ scattering\ states $\Psi ^{\pm }$
are solutions of the relativistic Schr\"{o}dinger equation 
\begin{equation}
H\Psi _{\mathbf{p}_{1}\mathbf{p}_{2}}^{\pm }=\left[ E_{1}\left( \mathbf{p}%
_{1}\right) +E_{2}\left( \mathbf{p}_{2}\right) \right] \Psi _{\mathbf{p}_{1}%
\mathbf{p}_{2}}^{\pm }.
\end{equation}%
Although not needed here, we note that the states $\Psi _{\mathbf{p}_{1}%
\mathbf{p}_{2}}^{+}$ and $\Psi _{\mathbf{p}_{1}\mathbf{p}_{2}}^{-}$ are the
solutions of the Lippmann-Schwinger equation 
\begin{eqnarray}
\Psi _{\mathbf{p}_{1}\mathbf{p}_{2}}^{\pm } &=&\left\vert \mathbf{p}_{1}%
\mathbf{p}_{2}\right\rangle +\frac{1}{E_{1}\left( \mathbf{p}_{1}\right)
+E_{2}\left( \mathbf{p}_{2}\right) -H_{0}\pm i0_{+}}H^{\prime }\Psi _{%
\mathbf{p}_{1}\mathbf{p}_{2}}^{\pm } \\
&=&\left\vert \mathbf{p}_{1}\mathbf{p}_{2}\right\rangle +\frac{1}{%
E_{1}\left( \mathbf{p}_{1}\right) +E_{2}\left( \mathbf{p}_{2}\right) -H\pm
i0_{+}}H^{\prime }\left\vert \mathbf{p}_{1}\mathbf{p}_{2}\right\rangle . 
\notag
\end{eqnarray}%
where $H^{\prime }$ is the interaction Hamiltonian 
\begin{equation}
H^{\prime }=H-H_{0}
\end{equation}%
while $\left\vert \mathbf{p}_{1}\mathbf{p}_{2}\right\rangle $\ is a solution
of 
\begin{equation}
H_{0}\left\vert \mathbf{p}_{1}\mathbf{p}_{2}\right\rangle =\left[
E_{1}\left( \mathbf{p}_{1}\right) +E_{2}\left( \mathbf{p}_{2}\right) \right]
\left\vert \mathbf{p}_{1}\mathbf{p}_{2}\right\rangle .
\end{equation}%
The scattering operator $S$\ is defined by \cite{Ne82, Ro65} 
\begin{equation}
\Psi _{\mathbf{p}_{1}\mathbf{p}_{2}}^{+}=\hat{S}\Psi _{\mathbf{p}_{1}\mathbf{%
p}_{2}}^{-}  \label{S_1}
\end{equation}%
and the S-matrix elements are accordingly determined by 
\begin{equation}
S\left( \mathbf{p}_{1}^{\prime },\mathbf{p}_{2}^{\prime };\mathbf{p}_{1},%
\mathbf{p}_{2}\right) =\left\langle \Psi _{\mathbf{p}_{1}^{\prime }\mathbf{p}%
_{2}^{\prime }}^{+}\right\vert \left. \hat{S}\Psi _{\mathbf{p}_{1}\mathbf{p}%
_{2}}^{+}\right\rangle =\left\langle \Psi _{\mathbf{p}_{1}^{\prime }\mathbf{p%
}_{2}^{\prime }}^{-}\right\vert \left. \Psi _{\mathbf{p}_{1}\mathbf{p}%
_{2}}^{+}\right\rangle  \label{S_2}
\end{equation}%
It has been proved (e.g. section 6 of \cite{FS64} ) that the
Bakamjian-Thomas construction guarantees the Poincar\'{e} invariance of the
operator $S$.

In \cite{FS64} has been shown that 
\begin{equation}
S\left( \mathbf{p}_{1}^{\prime },\mathbf{p}_{2}^{\prime };\mathbf{p}_{1},%
\mathbf{p}_{2}\right) =\left[ J\left( \mathbf{p}_{1}^{\prime },\mathbf{p}%
_{2}^{\prime }\right) J\left( \mathbf{p}_{1},\mathbf{p}_{2}\right) \right]
^{-\frac{1}{2}}\delta ^{3}\left( \mathbf{P}^{\prime }-\mathbf{P}\right)
\left\langle \Phi _{\mathbf{k}^{\prime }}^{-}\right\vert \left. \Phi _{%
\mathbf{k}}^{+}\right\rangle  \label{S_3}
\end{equation}%
where $\Phi _{\mathbf{k}}^{\pm }$ denote the \textquotedblleft
in\textquotedblright\ and \textquotedblleft out\textquotedblright\
eigenstate of the mass operator $\hat{M}$ , with asymptotic relative
momentum $\mathbf{k}$%
\begin{equation}
\hat{M}\Phi _{\mathbf{k}}^{\pm }=w\Phi _{\mathbf{k}}^{\pm }.  \label{m_1}
\end{equation}%
where the c.m. energy $w$\ is given by 
\begin{equation}
w=w\left( \mathbf{k}\right) =\sqrt{m_{1}^{2}+\mathbf{k}^{2}}+\sqrt{m_{2}^{2}+%
\mathbf{k}^{2}}
\end{equation}%
More precisely, the states $\Phi _{\mathbf{k}}^{\pm }$\ are the solutions of
the Lippmann-Schwinger equation 
\begin{eqnarray}
\Phi _{\mathbf{k}}^{\pm } &=&\left\vert \mathbf{k}\right\rangle +\frac{1}{%
w\left( \mathbf{k}\right) -\hat{M}_{0}\pm i0_{+}}\hat{V}\Phi _{\mathbf{k}%
}^{\pm } \\
&=&\left\vert \mathbf{k}\right\rangle +\frac{1}{w\left( \mathbf{k}\right) -%
\hat{M}\pm i0_{+}}\hat{V}\left\vert \mathbf{k}\right\rangle
\end{eqnarray}%
where $\left\vert \mathbf{k}\right\rangle $ is a solution of 
\begin{equation}
\hat{M}_{0}\left\vert \mathbf{k}\right\rangle =w\left\vert \mathbf{k}%
\right\rangle
\end{equation}

For two particles with equal masses $m_{1}=m_{2}=m$ the equation (\ref{m_1})
simplifies to 
\begin{equation}
(2\sqrt{m^{2}+\mathbf{\hat{k}}^{2}}+\hat{V})\Phi _{\mathbf{k}}^{\pm }=2\sqrt{%
m^{2}+\mathbf{k}^{2}}\Phi _{\mathbf{k}}^{\pm }  \label{m_2}
\end{equation}%
Squaring both sides and making some rearrangement, equation (\ref{m_2}) can
put in the form \cite{CPS75, CP82, GLC86} 
\begin{equation}
\left( \frac{\mathbf{\hat{k}}^{2}}{m}+\mathcal{\hat{V}}\right) \Phi _{%
\mathbf{k}}^{\pm }=\mathcal{E}\Phi _{\mathbf{k}}^{\pm }  \label{m_3}
\end{equation}%
with $\mathcal{E}=\mathbf{k}^{2}/m$ and 
\begin{equation}
\mathcal{\hat{V}}=\frac{1}{2m}\left\{ \hat{V},\sqrt{m^{2}+\mathbf{\hat{k}}%
^{2}}\right\} +\frac{\hat{V}^{2}}{4m}.
\end{equation}%
Equation (\ref{m_3}) is identical in structure to a non-relativistic Schr%
\"{o}dinger equation.

We can define $S$\ matrix related to (\ref{m_3}) 
\begin{equation}
\Phi _{\mathbf{k}}^{+}=\mathcal{S}\Phi _{\mathbf{k}}^{-}.  \label{s_1}
\end{equation}%
Note, this $S$\ matrix is different from the previous one. The principal
difference between the scattering operators in (\ref{S_1}) and (\ref{s_1})
is that while the former commutes with generators $H$, $\mathbf{P}$, $%
\mathbf{J}$ and $\mathbf{K}$ of the Poincar\'{e} group $\mathcal{P}$, the
latter commutes with generators $\mathbf{\hat{l}}$ of a \ group being
isomorphic to $SO(3)$.

According to (\ref{s_1})%
\begin{equation}
\left\langle \Phi _{\mathbf{k}^{\prime }}^{-}\right\vert \left. \Phi _{%
\mathbf{k}}^{+}\right\rangle =\left\langle \Phi _{\mathbf{k}^{\prime
}}^{+}\right\vert \left. \mathcal{S}\Phi _{\mathbf{k}}^{+}\right\rangle =%
\mathcal{S}\left( \mathbf{k}^{\prime },\mathbf{k}\right)  \label{s_2}
\end{equation}%
Separating from $\mathcal{S}\left( \mathbf{k}^{\prime },\mathbf{k}\right) $\
the non-interacting part, it is customary to write 
\begin{equation}
\mathcal{S}\left( \mathbf{k}^{\prime },\mathbf{k}\right) =\delta ^{3}\left( 
\mathbf{k}^{\prime }-\mathbf{k}\right) -2\pi i\delta \left( \mathcal{E}%
^{\prime }-\mathcal{E}\right) \mathcal{T}\left( \mathbf{k}^{\prime },\mathbf{%
k}\right) .  \label{s_3}
\end{equation}%
where $\mathcal{T}\left( \mathbf{k}^{\prime },\mathbf{k}\right) $\ is called 
$T$\ matrix or transition amplitude. Since $\mathcal{V}$\ is rotationally
invariant the transition amplitude $\mathcal{T}\left( \mathbf{k}^{\prime },%
\mathbf{k}\right) $ may be a function of $k\equiv \left\vert \mathbf{k}%
\right\vert $\ and $\cos \theta =\mathbf{n}^{\prime }\cdot \mathbf{n}$\
only, where $\mathbf{n}^{\prime }=\mathbf{k}^{\prime }/k^{\prime }$and$\ 
\mathbf{n=k}/k$. It related to the c.m. scattering amplitude $f\left( \theta
\right) $\ by the equation \cite{Sak85} 
\begin{equation}
\mathcal{T}\left( \mathbf{k}^{\prime },\mathbf{k}\right) =-\frac{1}{2\pi
^{2}m}f\left( \theta \right)
\end{equation}%
where $\theta $ is the c.m. scattering angle.

Inserting these relations into Eq.(\ref{S_3}) and using the identities 
\begin{equation}
\delta ^{3}\left( \mathbf{P}^{\prime }-\mathbf{P}\right) \delta \left( 
\mathcal{E}^{\prime }-\mathcal{E}\right) =\frac{2m}{E}\delta ^{3}\left( 
\mathbf{P}^{\prime }-\mathbf{P}\right) \delta \left( E^{\prime }-E\right)
\end{equation}%
and \cite{FS64} 
\begin{equation}
\delta ^{3}\left( \mathbf{P}^{\prime }-\mathbf{P}\right) \delta ^{3}\left( 
\mathbf{k}^{\prime }-\mathbf{k}\right) =J\left( \mathbf{p}_{1},\mathbf{p}%
_{2}\right) \delta ^{3}\left( \mathbf{p}_{1}^{\prime }-\mathbf{p}_{1}\right)
\delta ^{3}\left( \mathbf{p}_{2}^{\prime }-\mathbf{p}_{2}\right)
\end{equation}%
we find that 
\begin{equation}
S\left( \mathbf{p}_{1}^{\prime },\mathbf{p}_{2}^{\prime };\mathbf{p}_{1},%
\mathbf{p}_{2}\right) =\delta ^{3}\left( \mathbf{p}_{1}^{\prime }-\mathbf{p}%
_{1}\right) \delta ^{3}\left( \mathbf{p}_{2}^{\prime }-\mathbf{p}_{2}\right)
-2\pi i\delta ^{4}\left( p_{1}^{\prime }+p_{2}^{\prime }-p_{1}-p_{2}\right) 
\frac{\mathcal{M}\left( p_{1},p_{2};p_{1}^{\prime },p_{2}^{\prime }\right) }{%
\left( 2E_{1}\right) ^{1/2}\left( 2E_{2}\right) ^{1/2}\left( 2E_{1}^{\prime
}\right) ^{1/2}\left( 2E_{2}^{\prime }\right) ^{1/2}}
\end{equation}%
where 
\begin{equation}
\ \mathcal{M}=-\sqrt{s}f\left( \theta \right) /\pi ^{2}
\end{equation}%
is an invariant amplitude. Thus, in order to determine $S\left( \mathbf{p}%
_{1}^{\prime },\mathbf{p}_{2}^{\prime };\mathbf{p}_{1},\mathbf{p}_{2}\right) 
$, it is sufficient to know the c.m. scattering amplitude $f\left( \theta
\right) $ \ 
\begin{equation}
f\left( \theta \right) =\frac{1}{2ik}\sum_{l=0}^{\infty }\left( 2l+1\right)
\left( \mathcal{S}_{l}-1\right) P_{l}\left( \cos \theta \right)  \label{par}
\end{equation}%
where $P_{l}$ are Legendre polynomials and $\mathcal{S}_{l}$\ is the $S$\
matrix element for angular momentum $l$.

The two-body cross section is given by 
\begin{equation}
\frac{d\sigma }{dt}=\frac{\pi ^{5}\left| \ \mathcal{M}\right| ^{2}}{\sqrt{%
\left( p_{1}p_{2}\right) ^{2}-m^{4}}}  \label{cross_1}
\end{equation}
where $t$ is the momentum transfer squared, i.e., 
\begin{equation}
t=\left( p_{1}^{\prime }-p_{1}\right) ^{2}.
\end{equation}
It may be written as 
\begin{equation}
\frac{d\sigma }{d\Omega }=\left| f\left( \theta \right) \right| ^{2}
\label{cross_2}
\end{equation}

It seems reasonable to assume that there might be a relativistic interacting
system that has a non-compact group $G$ as dynamical symmetry group in the
sense that 
\begin{equation}
\frac{\mathbf{\hat{k}}^{2}}{m}+\mathcal{\hat{V}}=f\left( C\right) .
\end{equation}%
where $C$\ is the Casimir operator of $G$. If that is the case, then the $S$
matrix for systems under consideration is constrained to satisfy \cite{Ke98} 
\begin{equation}
\mathcal{S}T^{\chi }\left( g\right) =T^{\tilde{\chi}}\left( g\right) 
\mathcal{S}\;,\qquad \text{for\ all}\;g\in G  \label{SU_1}
\end{equation}%
or equivalently, 
\begin{equation}
\mathcal{S}dT^{\chi }\left( a\right) =dT^{\tilde{\chi}}\left( a\right) 
\mathcal{S},\qquad \text{for\ all}\;a\in \mathfrak{g}  \label{SU_2}
\end{equation}%
where $T^{\chi }$ and $T^{\tilde{\chi}}$ are Weyl-equivalent representations
of $G$ specified by labels $\chi $ and $\tilde{\chi}$, while $dT^{\chi }$
and $dT^{\tilde{\chi}}$ are the corresponding representations of the algebra 
$\mathfrak{g}$ of $G$. ( The representations $T^{\chi }$ and $T^{\widetilde{%
\chi }}$ have the same Casimir eigenvalues. Such representations are called
Weyl equivalent.) Eqs (\ref{SU_1}) and (\ref{SU_2}) have much restriction
power and are used in deriving the $S$ matrix \cite{Ke05}.

In order to avoid misunderstanding, we make a few comments on the equation (%
\ref{SU_1}) or (\ref{SU_2}). To start with, it should be pointed out that we
have in the subspace of scattering states two complete orthonormal systems, $%
\left\{ \Phi ^{+}\right\} $ and $\left\{ \Phi ^{-}\right\} $. The state
vectors $\Phi ^{-}$\ transform according to the representation $T^{\chi
}\left( g\right) $, while the state vectors$\ \Phi ^{+}$\ transform
according to $T^{\tilde{\chi}}\left( g\right) $. Since by definition the
operator $\mathcal{S}$\ maps each $\Phi ^{-}$\ on the corresponding $\ \Phi
^{+},$ then $T^{\tilde{\chi}}\left( g\right) \Phi ^{+}=\mathcal{S}T^{\chi
}\left( g\right) \Phi ^{-}$\ so that $T^{\tilde{\chi}}\left( g\right) 
\mathcal{S}\Phi ^{-}=\mathcal{S}T^{\chi }\left( g\right) \Phi ^{-}$. This
means that $\mathcal{S}$ must satisfy the equation (\ref{SU_1}). Moreover,
if $\mathcal{S}$ intertwines representations $T^{\chi }$ and $T^{\tilde{\chi}%
}$ of the Lie group $G$, it also intertwines the representations $dT^{\chi }$
and $dT^{\tilde{\chi}}$ of the Lie algebra $\mathfrak{g}.$

Finally, we would like to emphasize that,\ in general, $\mathbf{\hat{k}}%
^{2}/m+\mathcal{\hat{V}}$\ is some rotationally invariant operator function
of $\mathbf{\hat{k}}$\textbf{\ }and $\mathbf{\hat{\rho}}$. So the
geometrical invariance algebra of (\ref{m_3})\ is the algebra generated by $%
\mathbf{\hat{l}}$. In other words, the group $G$ has a subgroup being
isomorphic to $SO(3)$. Since the representations $T^{\chi }$ and $T^{%
\widetilde{\chi }}$ are identical for compact subgroups of $G$ \cite{KS71,
KS80}, it follows from (\ref{SU_1}) and (\ref{SU_2}) that%
\begin{equation}
\left[ T^{\chi }\left( g\right) ,\mathcal{S}\right] =0,\quad \text{if \ \ }%
g\in SO(3).
\end{equation}%
or 
\begin{equation}
\left[ dT^{\chi }\left( a\right) ,\mathcal{S}\right] =0,\quad \text{if \ \ }%
a\in \mathfrak{so(}3\mathfrak{)}.
\end{equation}%
as it should be.

Let us apply this construction to scattering systems that have $SO(3,1)$ as
dynamical symmetry group, i.e. 
\begin{equation}
\frac{\mathbf{\hat{k}}^{2}}{m}+\mathcal{\hat{V}}=f\left( C_{1}\right) .
\label{V_C}
\end{equation}%
We first note that \cite{San65, San67} the internal angular momentum
operator $\mathbf{\hat{l}}$\ and the operator $\mathbf{\hat{N}}$\ defined by 
\begin{equation}
\mathbf{\hat{N}=}\frac{1}{2\sqrt{\mathbf{\hat{k}}^{2}}}\left[ \mathbf{\hat{k}%
\times \hat{l}-\hat{l}\times \hat{k}}\right] 
\end{equation}%
span the Lie algebra of the Lorentz group $SO(3,1)$%
\begin{equation}
\left[ \hat{l}_{i},\hat{l}_{j}\right] =i\epsilon _{ijk}\hat{l}_{k}\text{ , \
\ }\left[ \hat{l}_{i},N_{j}\right] =i\epsilon _{ijk}N_{k}\text{ , \ \ \ }%
\left[ N_{i},N_{j}\right] =-i\epsilon _{ijk}\hat{l}_{k}  \label{L_A}
\end{equation}%
(The dynamical algebra (\ref{L_A}) should not be confused with Lorentz
subalgebra (\ref{P_2}) of the Poincar\'{e} group.) These commutation
relations can be easily calculated by making use of Eqs (\ref{ccr_1}) and 
\begin{equation}
\left[ \hat{l}_{i},\hat{k}_{j}\right] =i\epsilon _{ijk}\hat{k}_{k}
\label{vect}
\end{equation}%
Then, the $S$ matrices for such systems can be obtained from equation (\ref%
{SU_1}) or (\ref{SU_2}). To this end, a few facts from representation theory
of the group $SO(3,1)$ are useful.

The unitary irreducible representations of $SO(3,1)$ are known to form three
series: principal, supplementary and discrete. It is also known that only
the principal series describes the scattering states. The principal series
of $SO(3,1)$ are characterized by the pair $\chi =\left( \tau ,\lambda
\right) ,$ where $\lambda =0,\pm \frac{1}{2},\pm 1,\ldots $ , while $-\infty
<\tau <\infty .$ The representations specified by labels $\chi =\left( \tau ,%
\text{ }\lambda \right) $ and $\tilde{\chi}=\left( -\tau ,-\lambda \right) $
are Weyl equivalent. In every UIR of principal series of $SO(3,1)$ the
Casimir invariants $C_{1}$ and $C_{2}$ 
\begin{equation}
C_{1}=\mathbf{J}^{2}-\mathbf{N}^{2},\text{ \ \ }C_{2}=\mathbf{J\cdot N}
\end{equation}%
become equal to a multiple of the identity operator $I$ 
\begin{equation}
C_{1}=-\left( \lambda ^{2}+\tau ^{2}+1\right) I,\text{ \ \ \ \ }%
C_{2}=\lambda \tau I.
\end{equation}%
where\ $\mathbf{J}$ and $\mathbf{N}$ are the generators of rotations, and
pure Lorentz transformations, respectively.

It is also worth noticing \cite{San67} \ that the second Casimir invariant $%
C_{2}$\ is identically zero for the above realization of $SO(3,1)$%
\begin{equation}
C_{2}=\frac{1}{2\sqrt{\mathbf{\hat{k}}^{2}}}\left[ \mathbf{\hat{l}\cdot }%
\left( \mathbf{\hat{k}\times \hat{l}}\right) \mathbf{-\hat{l}\cdot }\left( 
\mathbf{\hat{l}\times \hat{k}}\right) \right] =0
\end{equation}
Consequently, the relevant unitary representations will be the principal
series representation $\left( \tau ,0\right) $. It is worthwhile to point
out that the second label, $\lambda $, of the $\left( \tau ,\lambda \right) $
irrep is connected with the helicities of particles; that is why we have the 
$\left( \tau ,0\right) $ irrep for spinless particles.

The representations specified by $\chi =\left( \tau ,0\right) $ can be
realized in the Hilbert space spanned by the eigenvectors $\left\vert \tau
0;l\mu \right\rangle $\ of $\mathbf{\hat{l}}^{2}$\ and $\hat{l}_{3}$.\ The
operators $\hat{l}_{i},\hat{N}_{i}$ are then defined by 
\begin{eqnarray*}
\hat{l}_{3}^{\chi }\left\vert l\mu \right\rangle &=&\mu \left\vert l\mu
\right\rangle \\
\hat{l}_{\pm }^{\chi }\left\vert l\mu \right\rangle &=&\left[ \left( l\mp
\mu \right) \left( l\pm \mu +1\right) \right] ^{\frac{1}{2}}\left\vert l,\mu
\pm 1\right\rangle \\
\hat{N}_{3}^{\chi }\left\vert l\mu \right\rangle &=&i\left( -1+i\tau
-l\right) a_{l+1,\mu }\left\vert l+1,\mu \right\rangle +i\left( i\tau
+l\right) a_{l,\mu }\left\vert l-1,\mu \right\rangle \\
\hat{N}_{\pm }^{\chi }\left\vert l\mu \right\rangle &=&\pm i\left( 1-i\tau
+l\right) b_{l+1,\pm \mu +1}\left\vert l+1,\mu \pm 1\right\rangle \pm
i\left( i\tau +l\right) b_{l,\mp \mu }\left\vert l-1,\mu \pm 1\right\rangle
\end{eqnarray*}%
where $\hat{N}_{\pm }=\hat{N}_{1}\pm i\hat{N}_{2},$ \ $\hat{N}_{\pm }=\hat{N}%
_{1}\pm i\hat{N}_{2},$\ $\left\vert l\mu \right\rangle \equiv \left\vert
\tau 0;l\mu \right\rangle $ and 
\begin{equation}
a_{l,\mu }=\sqrt{\frac{\left( l+\mu \right) \left( l-\mu \right) }{\left(
2l+1\right) \left( 2l-1\right) }},\text{ \ \ }b_{l,\mu }=\sqrt{\frac{\left(
l+\mu \right) \left( l+\mu -1\right) }{\left( 2l+1\right) \left( 2l-1\right) 
}}
\end{equation}

We can now evaluate the $S$\ matrix from (\ref{SU_2}). To do this let us
write equation (\ref{SU_2}) for generators $\hat{l}_{3},\hat{l}_{\pm }$ and $%
\hat{N}_{3}$%
\begin{eqnarray}
\mathcal{S}\hat{l}_{3}^{\chi } &=&\hat{l}_{3}^{\tilde{\chi}}\mathcal{S}
\label{l_3} \\
\mathcal{S}\hat{l}_{\pm }^{\chi } &=&\hat{l}_{\pm }^{\tilde{\chi}}\mathcal{S}
\label{l_pm} \\
\mathcal{S}\hat{N}_{3}^{\chi } &=&\hat{N}_{3}^{\tilde{\chi}}\mathcal{S}
\label{N_3}
\end{eqnarray}%
Applying both sides of equations (\ref{l_3}) and (\ref{l_pm}) to the basis
vector $\left\vert l\mu \right\rangle $ we find that the $\mathcal{S}$%
-matrix in the angular momentum representation is diagonal and its matrix
elements are independent of $\mu $, i.e., 
\begin{equation}
\mathcal{S}\left\vert l\mu \right\rangle =\mathcal{S}_{l}\left\vert l\mu
\right\rangle .
\end{equation}%
(Observe that the operator $\mathcal{S}$\ commutes with all $J_{i}^{\chi }$%
's, as expected.) The value of its diagonal elements can be defined by using
of (\ref{l_3}). As a result we obtain the recurrence relation 
\begin{equation}
\left( 1-i\tau +l\right) \mathcal{S}_{l+1}=\left( 1+i\tau +l\right) \mathcal{%
S}_{l},\text{ \ \ }\left( -i\tau +l\right) \mathcal{S}_{l}=\left( i\tau
+l\right) \mathcal{S}_{l-1}
\end{equation}%
which implies that 
\begin{equation}
\mathcal{S}_{l}=\frac{\Gamma \left( 1+i\tau +l\right) }{\Gamma \left(
1-i\tau +l\right) }  \label{P_A}
\end{equation}%
Inserting this into Eq.(\ref{par}) we obtain 
\begin{equation}
f\left( \theta \right) =\frac{1}{2ik}\frac{\Gamma \left( 1+i\tau \right) }{%
\Gamma \left( -i\tau \right) }\frac{1}{\sin ^{2}\dfrac{\theta }{2}}\exp %
\left[ -i\tau \ln \left( \sin ^{2}\dfrac{\theta }{2}\right) \right] ,\text{
\ }\theta \neq 0
\end{equation}%
where the momentum-dependent parameter $\tau $\ is determined by the
relation (\ref{V_C}). (For (\ref{P_A}) the expansion diverges as a function,
but it converge as a distribution \cite{Ta74}.) For example, in analogy to
the non-relativistic Coulomb interaction, we can propose 
\begin{equation}
\frac{\mathbf{\hat{k}}^{2}}{m}+\mathcal{\hat{V}}=-\frac{\alpha ^{2}m}{%
4\left( C_{1}+1\right) }
\end{equation}%
where $\alpha $\ denotes the strength of interaction. This means that 
\begin{equation}
\left( \frac{\mathbf{\hat{k}}^{2}}{m}+\mathcal{\hat{V}}\right) \Phi _{%
\mathbf{k}}^{\pm }=-\left( \frac{\alpha ^{2}m}{4\left( C_{1}+1\right) }%
\right) \Phi _{\mathbf{k}}^{\pm }
\end{equation}%
So, taking into account the Eqs. (\ref{m_3}) and 
\begin{equation}
C_{1}\Phi _{\mathbf{k}}^{\pm }=-\left( \tau ^{2}+1\right) \Phi _{\mathbf{k}%
}^{\pm }
\end{equation}%
we have 
\begin{equation}
\tau =\frac{\alpha m}{2k}.
\end{equation}%
It then follows that (in ordinary units) 
\begin{equation}
f_{c}\left( \theta \right) =\frac{\alpha }{mv^{2}\sin ^{2}\dfrac{\theta }{2}}%
\left( 1-\beta ^{2}\right) \exp \left[ -i\tau \ln \left( \sin ^{2}\frac{%
\theta }{2}\right) +i\pi +2i\eta \right] ,\text{ \ }\theta \neq 0
\end{equation}%
with $\tau =\dfrac{\alpha m}{2\hbar k}=\dfrac{\alpha }{\hbar v}\left(
1-\beta ^{2}\right) ^{1/2}$.$\ $Here$\ \beta =v/c,\,\eta =\arg \Gamma \left(
1+i\tau \right) $ and $v$\ is the relative velocity of the particles.

If two spinless particles are identical the indistinguishability of them
leads to the c.m. scattering amplitude $f_{c}^{s}\left( \theta \right) $ of
the form 
\begin{equation}
f_{c}^{s}\left( \theta \right) =f_{c}\left( \theta \right) +f_{c}\left( \pi
-\theta \right)
\end{equation}
This results in the differential cross section, 
\begin{eqnarray}
\frac{d\sigma }{d\Omega } &=&\left| f_{c}\left( \theta \right) +f_{c}\left(
\pi -\theta \right) \right| ^{2}  \notag \\
&=&\left( \dfrac{\alpha }{mv^{2}}\right) ^{2}\left\{ \dfrac{1}{\sin ^{4}%
\dfrac{\theta }{2}}+\frac{1}{\cos ^{4}\dfrac{\theta }{2}}+\frac{2}{\sin ^{2}%
\dfrac{\theta }{2}\cos ^{2}\dfrac{\theta }{2}}\cos \left[ \dfrac{\alpha }{%
\hbar v}\left( 1-\beta ^{2}\right) ^{1/2}\ln \tan ^{2}\frac{\theta }{2}%
\right] \right\} \left( 1-\beta ^{2}\right) ^{2}.
\end{eqnarray}
If we take the non-relativistic limit $\beta \rightarrow 0$, we gain the
Mott formula \cite{JCJ75} for the Coulomb scattering of two identical
spinless bosons.

\section{Conclusions and outlook}

In this paper we developed an algebraic technique for building two-body
relativistic models in the framework of the direct-interaction theories,
i.e., theories in which there are no external fields. We have demonstrated
how the algebraic technique \cite{Ke98}, originally conceived for
non-relativistic scattering, can be generalised for the construction of
relativistic scattering matrix. Crucial in all the developments was the
assumption that the mass operator $M$ for given scattering system is related
to the Casimir operator of some non-compact group $G$. The results described
in this paper could be extended in several ways. One of these would be their
use for the scattering models with spins. Also, our analysis has been
restricted to two-body systems. It would be interesting to generalise the
technique discussed in this paper to the study of scattering problems for
many-body systems. These and other extensions will be studied in subsequent
papers.

\end{document}